\documentclass[lettersize,journal]{IEEEtran}

\usepackage{amsmath,amsfonts}
\usepackage{algorithmic}
\usepackage{array}
\usepackage[caption=false,font=normalsize,labelfont=sf,textfont=sf]{subfig}
\usepackage{textcomp}
\usepackage{stfloats}
\usepackage{url}
\usepackage{verbatim}
\usepackage{graphicx}

\usepackage{cite}
\usepackage{amssymb}
\usepackage{xcolor}
\usepackage{booktabs}
\usepackage{multirow}
\usepackage{tabularx}
\usepackage{hyperref}
\usepackage{xurl}
\usepackage{float}

\def\BibTeX{{\rm B\kern-.05em{\sc i\kern-.025em b}\kern-.08em
    T\kern-.1667em\lower.7ex\hbox{E}\kern-.125emX}}
\usepackage{balance}
\begin{document}
\title{Video-Foley: Two-Stage Video-To-Sound Generation \\via Temporal Event Condition For Foley Sound}
\author{Junwon Lee, Jaekwon Im, Dabin Kim, \IEEEmembership{Graduate Student Member, IEEE}, Juhan Nam, \IEEEmembership{Member, IEEE}
\thanks{This work was supported by Institute of Information \& communications Technology Planning \& Evaluation (IITP) grant funded by the Korea government (MSIT) (No.RS-2019-II190075, Artificial Intelligence Graduate School Program (KAIST)) and the National Research Foundation of Korea (NRF) grant funded by the Korea government (MSIT) (No. RS-2023-00222383).}
\thanks{J. Lee and J. Nam are with Graduate School of AI, KAIST, Daejeon, Korea}
\thanks{J. Im, D. Kim, and J. Nam are with Graduate School of CT, KAIST, Daejeon, Korea}
}

\markboth{IEEE/ACM TRANSACTIONS ON AUDIO, SPEECH, AND LANGUAGE PROCESSING}%
{\MakeLowercase{\textit{Lee et al.}} Video-Foley: Two-Stage Video-To-Sound Generation via Temporal Event Condition For Foley Sound}



\maketitle

\newcommand{\jw}[1]{\textcolor{blue}{(#1)}}

\begin{abstract}
Foley sound synthesis is crucial for multimedia production, enhancing user experience by synchronizing audio and video both temporally and semantically. 
Recent studies on automating this labor-intensive process through video-to-sound generation face significant challenges. Systems lacking explicit temporal features suffer from poor alignment and controllability, while timestamp-based models require costly and subjective human annotation.
We propose \textbf{\textit{Video-Foley}}, a video-to-sound system using Root Mean Square (RMS) as an intuitive condition with semantic timbre prompts (audio or text). RMS, a frame-level intensity envelope closely related to audio semantics, acts as a temporal event feature to guide audio generation from video.
The annotation-free self-supervised learning framework consists of two stages, Video2RMS and RMS2Sound, incorporating mu-law scaled RMS discretization and RMS-ControlNet with a pretrained text-to-audio model.
Our extensive evaluation shows that Video-Foley achieves state-of-the-art performance in audio-visual alignment and controllability for sound timing, intensity, timbre, and nuance.
Source code, model weights and demos are available on our companion website\footnote{\url{https://jnwnlee.github.io/video-foley-demo}}.

\end{abstract}

\begin{IEEEkeywords}
Video-to-Sound, Video-to-Audio, Controllable Audio Generation, Multimodal Deep Learning.
\end{IEEEkeywords}

\section{Introduction}
\IEEEPARstart{F}{oley} is the process of designing and recording sound effects to enrich the auditory experience in film, television, video games, virtual reality, and other media. While video content contains various types of sounds—speech, music, and sound effects (SFX)—Foley specifically focuses on SFX, such as environmental and interaction-based sounds. These effects enhance the realism of visual content, compensating for audio details that are often unclear or absent during filming or production. This practice ensures that the audio aligns seamlessly with the visual narrative, capturing its semantics and temporal dynamics. 

However, accurately synchronizing sounds with the timing, intensity, timbre, and nuance of visual elements remains a labor-intensive task. Unlike visually irrelevant sounds, such as background music or off-screen speech, visually relevant SFX—like footsteps or a door slam—must be carefully aligned with their corresponding actions in the video \cite{regnet}. This synchronization challenge is further complicated by the distinction between foreground and background sounds. Foreground sounds, typically produced by main objects, are transient and event-driven, requiring fine-grained temporal control, while background sounds are more stationary and ambient \cite{varietysound,action2sound,dcase2024}. Although ambient sounds exhibit low temporal variation and can be synthesized from static inputs such as images or text \cite{clipsonic,audioldm}, foreground SFX demand precise timing and nuanced control, making their manual production highly labor-intensive. While using pre-recorded sound samples can eliminate the need for recording, it involves extensive database searches and precise synchronization. These challenges highlight the need for automation or assistance in Foley \cite{oh2023demand}.


\begin{figure}[t!]
    \centering
    \resizebox{\columnwidth}{!}{%
        \includegraphics{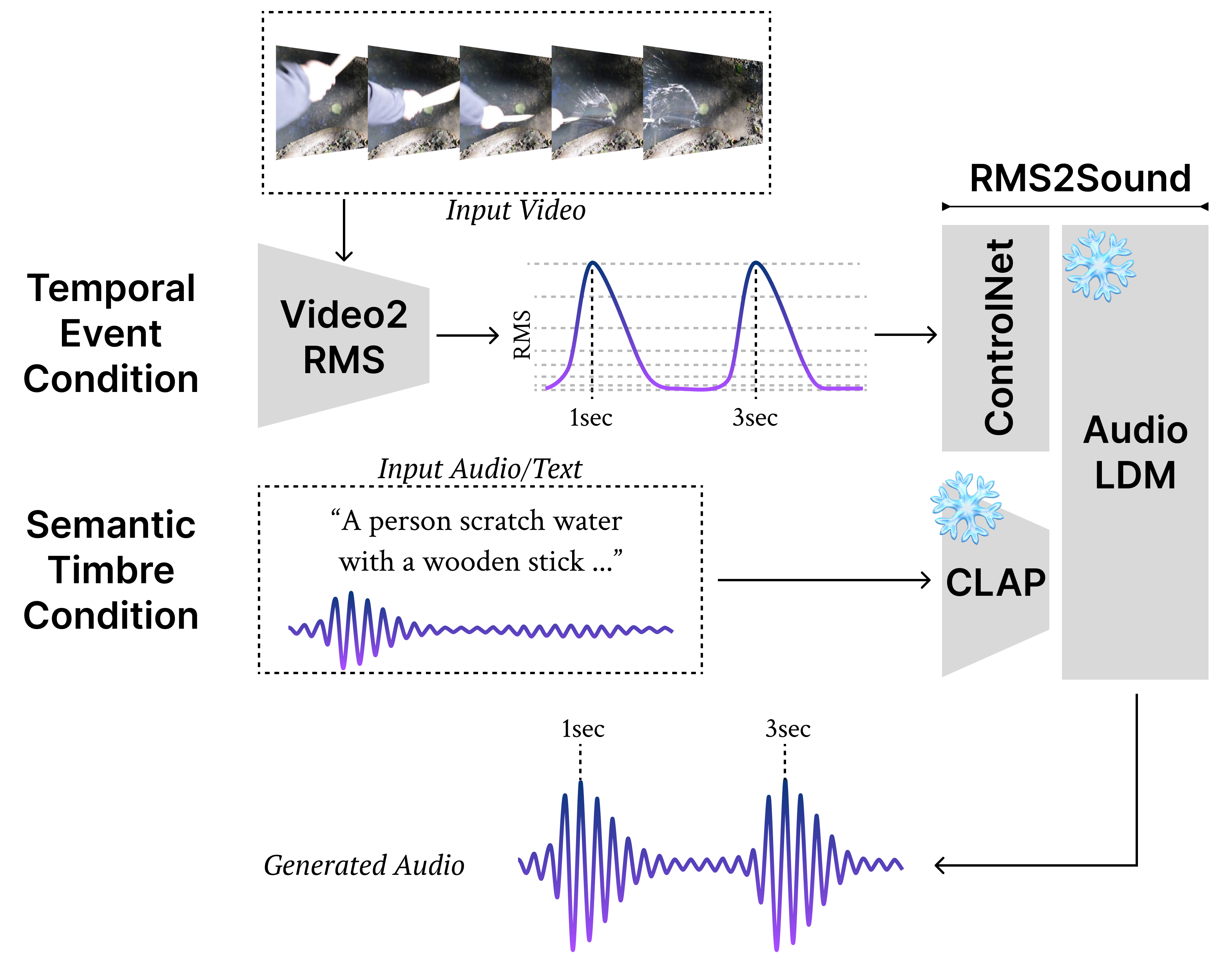}
    }
    \caption{Overall pipeline of the proposed model, a two-stage Video-to-Sound generation framework. Note that RMS can be extracted from audio waveform numerically. Video2RMS and RMS2Sound parts are trained separately.}
    \label{fig:model}
    \vspace{-0.3cm}
\end{figure}

Recent advances in generative AI have encouraged researchers to explore models that learn the cross-modal correspondence and synthesize audio content directly from video input. In video-to-sound generation, achieving both semantic and temporal synchronization between the two modalities is crucial. However, existing studies have not successfully accomplished this dual objectives. 
Early video-to-sound models aimed to generate audio from video by learning semantic correspondence using unsupervised training such as GAN \cite{regnet,specvqgan,foleygan}. However, they often struggled with poor temporal alignment and audio quality due to the lack of direct temporal supervision and low data quality.
More recent studies have explicitly incorporated temporal information into their models. One approach utilizes a holistic latent feature that captures both semantic and temporal information, employing techniques like contrastive audio-visual learning \cite{diff-foley} or knowledge distillation leveraging pretrained audio and vision embeddings \cite{maskvat}. However, low video framerate due to local temporal coherence and high computational costs has limited temporal alignment accuracy.
Other methods use sound event timestamps (e.g., onset, offset) as temporal features, combined with text prompts to guide audio generation \cite{syncfusion,sonicvisionlm}. They trained timestamp detection networks to classify each video frame via supervised learning. However, this method requires human annotation, which is costly and often ambiguous in defining precise time boundaries. Additionally, simply detecting the start and end points of sound events misses many important aspects of audio, such as the volume dynamics of a moving car, which are difficult to represent in text \cite{t-foley}.

We propose \textbf{\textit{Video-Foley}}, 
a two-stage model that leverages temporal event conditions for annotation-free training of highly synchronized Foley sound generation.
Rather than generating audio directly from video frames, our model first predicts a temporal feature as an intermediate representation, then generates audio from it.
At its core, we introduce the Root Mean Square (RMS) of audio content as a key temporal feature.
Defined as a frame-level energy feature calculated from audio waveforms, RMS captures not only the presence of sound events but also their intensity and temporal change, associated with subtle timbre and nuance~\cite{t-foley,mambafoley}. 
We propose to incorporate RMS as a target in the video encoding stage to ensure strong temporal and semantic audio-visual synchronization. Together with audio or text prompts, RMS serves as a control condition in the audio generation stage, enhancing controllability.
Our two-stage framework, illustrated in Fig. \ref{fig:model}, consists of Video2RMS and RMS2Sound. Video2RMS first predicts the RMS curve from video effectively using techniques such as label discretization and smoothing. Subsequently, RMS2Sound takes the RMS with an audio or text prompt to generate a temporally and semantically aligned audio waveform.
Inspired by ControlNet \cite{controlnet}, designed to add spatial conditioning (e.g., sketch) to pretrained image generators, our proposed RMS-ControlNet guides a frozen text-to-audio model \cite{audioldm} via an RMS curve.
The two modules are trained separately: Video2RMS trained on video-audio pairs (i.e., general video files), while RMS2Sound trained on audio-only data—both without any human annotations.
Through objective evaluation metrics and subjective human listening test, we demonstrate that Video-Foley achieves state-of-the-art performance in both temporal and semantic alignment on the Greatest Hits dataset~\cite{greatesthits}. Additionally, qualitative analysis and accompanying demo highlight its high controllability over timing, intensity, timbre, and nuance in the generated audio.

\section{Related Works}
In the early stages of automated Foley synthesis, parametric rule-based algorithms were used for constrained scenarios \cite{menexopoulos2023state}. For instance, simulated motion data were mapped to the parameters of a sound synthesis module \cite{obrien2001synthesizing}. More recent studies generate raw audio directly from video in an end-to-end manner, enabled by advances in deep learning. In this section, we review neural video-to-sound generation approaches as well as controllable audio generation with temporal conditioning, which is directly relevant to our proposed method.



\begin{figure}[t]
    \centering
    \resizebox{0.9\columnwidth}{!}{
        \includegraphics{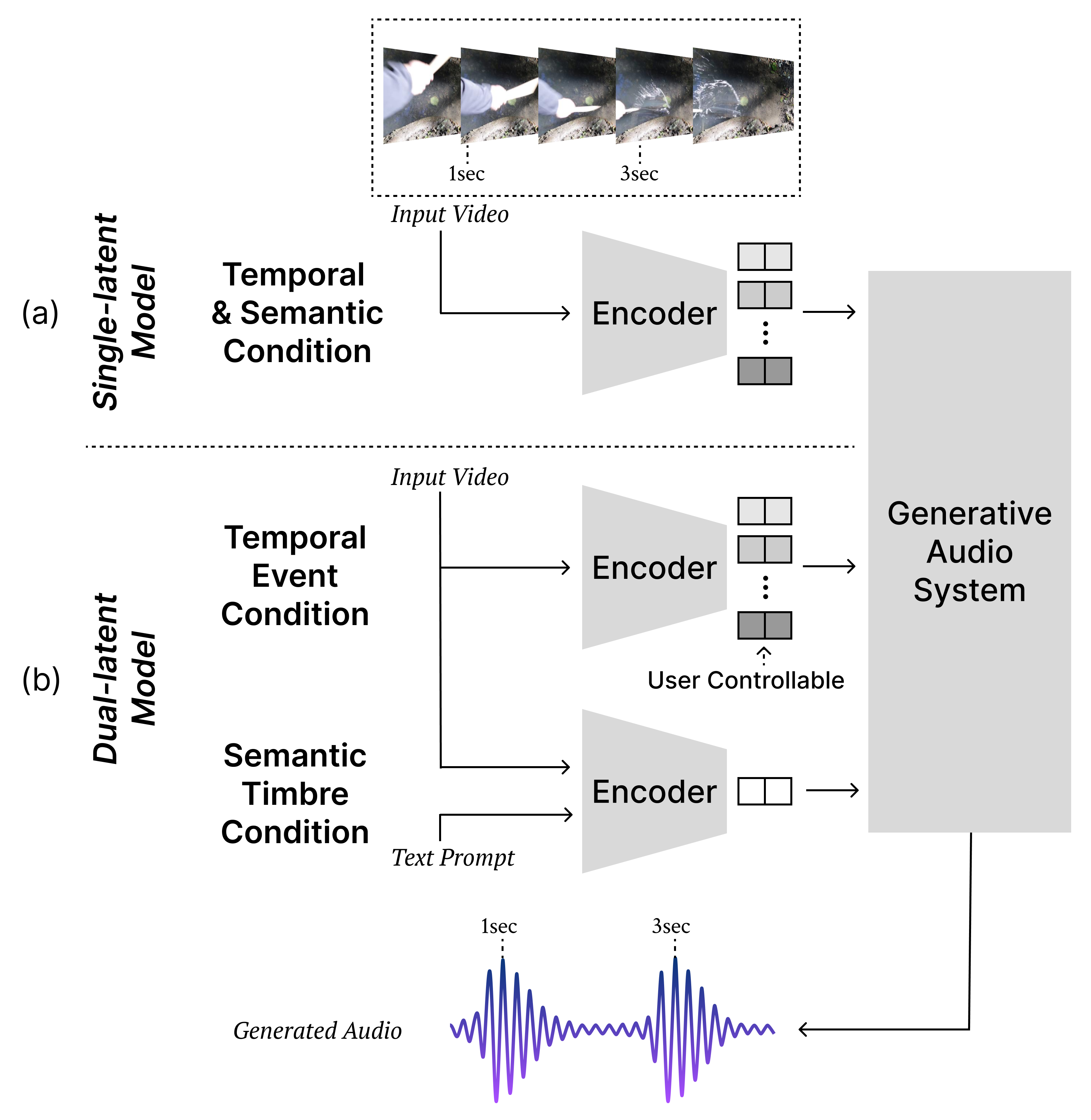}
    }
    \vspace{-0.2cm}
    \caption{Two model architecture types in neural video-to-sound generation: (a) single-latent model and (b) dual-latent model.}
    \label{fig:model_type}
    \vspace{-0.3cm}
\end{figure}

\subsection{Neural Video-to-Sound Generation}
\label{sec:v2s_gen}

In video-to-sound generation, achieving both semantic and temporal synchronization between the two modalities is crucial. Namely, the generated audio should have appropriate \textit{semantic content} (e.g., timbre, nuance, spatial attributes) and \textit{temporal content} (e.g., timing, intensity dynamics) that aligns with the video. There are two types of architecture for neural video-to-sound models, as shown in Fig. \ref{fig:model_type}: models with a single entangled latent space for both temporal and semantic information (single-latent model) and models with two separate latent spaces for temporal and semantic features (dual-latent model). 
In contrast to single-latent models, dual-latent models aim to provide user control over temporal or semantic information by using interpretable features or additional modalities, such as text prompts.

Despite advancements, existing studies have not fully accomplished this dual goal.
Early video-to-sound models, such as GAN-based methods \cite{regnet,specvqgan,foleygan}, aimed to generate audio from video input in an unsupervised manner (single-latent). These focused on learning semantic audio-visual correspondence from datasets of in-the-wild quality.
Subsequent works introduced controllable video-to-sound generation models that allow timbre adjustment via audio prompts \cite{varietysound} or audio-visual correlations \cite{condfoleygen} (dual-latent).    
Though they showed promising results, these approaches often suffered from temporal misalignment and low audio quality due to insufficient temporal guidance or low data quality.


Recent studies have explicitly incorporated temporal information into their models. For instance, Diff-Foley \cite{diff-foley} utilized a temporal-aware audio-visual joint embedding space trained through contrastive learning to condition a diffusion model for audio generation. 
MaskVAT \cite{maskvat} trained a transformer-based encoder using knowledge distillation, transferring sequential embeddings from a pretrained audio classifier to embeddings from a pretrained vision encoder.
However, low visual temporal resolution limited the accuracy of temporal alignment in these single-latent approaches. In case of Diff-Foley, a low video frame rate (4 fps) is unavoidable due to the high computational cost of the contrastive learning setting and the local temporal coherence of video frames (i.e., adjacent video frames are similar to each other). This often results in audio generation that is off-sync with video beyond the perceptual just noticeable difference ($\sim$50 ms) \cite{jnd}.
Other dual-latent approaches using temporal feature are discussed in the subsequent subsection.



\subsection{Neural Audio Generation with Temporal Event Condition} \label{ssec:con_audio_gen}

\begin{figure}[t]
    \centering

    \resizebox{0.55\columnwidth}{!}{%
        \includegraphics{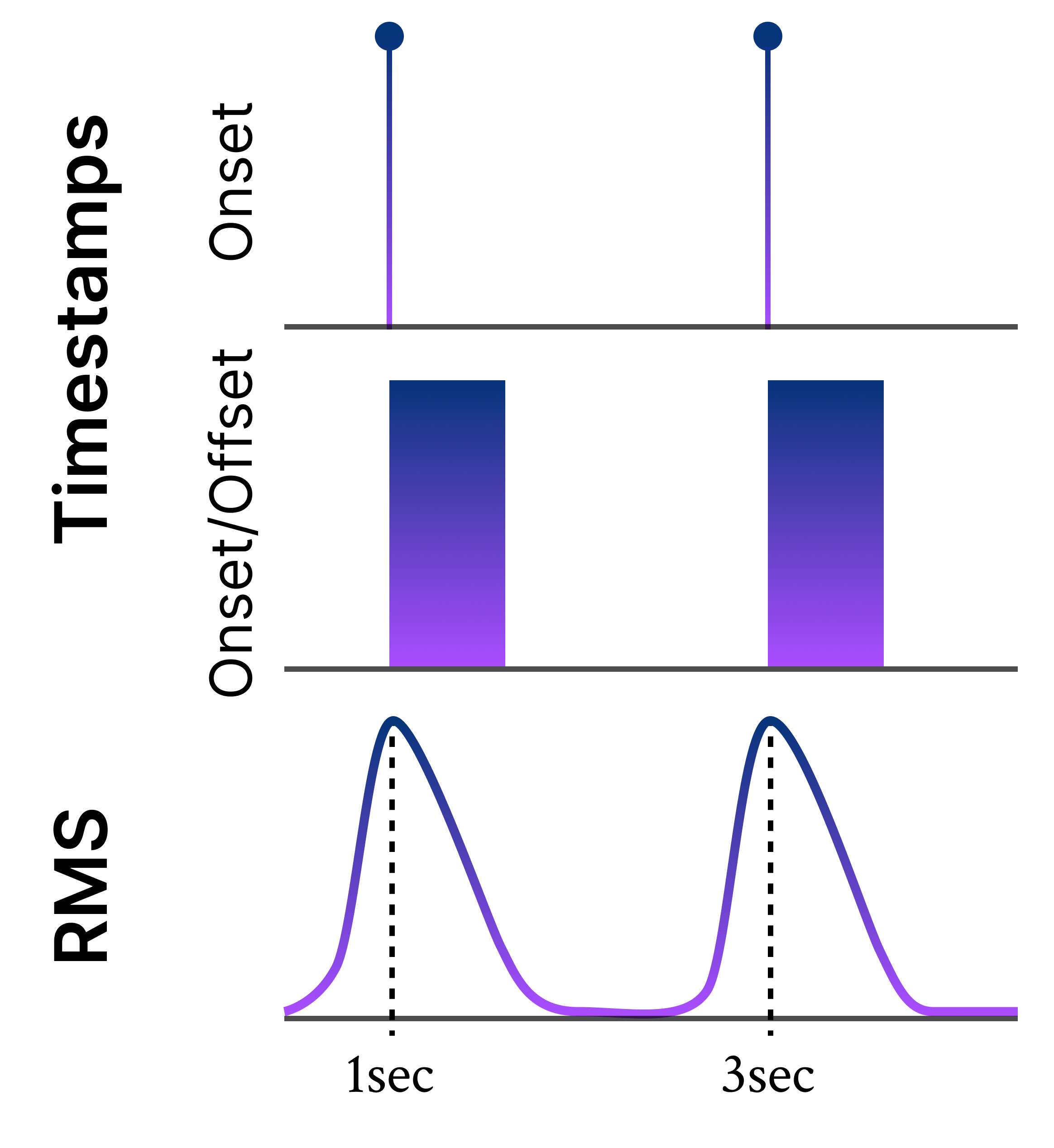}
    }
    \vspace{-0.3cm}
    \caption{Three types of temporal features}
    \label{fig:temporal_features}
    \vspace{-0.3cm}
\end{figure}

To bridge the gap between text/video and audio, interpretable temporal features have been introduced as temporal event conditions for audio generation. These features serve as intuitive guides for audio generation, much like a sketch guides image generation. Fig. \ref{fig:temporal_features} shows three different types of features. 
The first is timestamps, which define sound event boundaries using onset (the start of a sound) or a combination of onset and offset (the start and end). As a high-level feature, timestamps provide a simple yet effective way to represent transient sound events. 
Another temporal feature is RMS, a windowed root-mean-squared value of the audio waveform. RMS not only implicitly captures sound onsets and offsets but also reflects frame-level intensity dynamics, offering a more detailed temporal representation.

Various neural audio generation approaches have leveraged these temporal features to enable temporal event control. A straightforward approach is to train a model from scratch that directly takes these conditioning inputs. T-Foley \cite{t-foley} introduced a waveform domain diffusion model conditioned on both RMS and sound category text to guide audio generation. The results from T-Foley and other follow-up studies, such as MambaFoley~\cite{mambafoley}, demonstrated that RMS effectively controls the temporal characteristics while also influencing semantic elements such as timbre. Moreover, they showed that RMS can be easily manipulated by users through simple actions like voice or clap sounds.

Other methods have focused on enhancing the controllability of pretrained large-scale generative models by incorporating new input types, leveraging these models' performance and generalizability.
In the vision domain, ControlNet \cite{controlnet} introduces additional spatial conditioning controls—such as sketch, depth, and human pose—into a high-performance text-to-image latent diffusion model (LDM). Instead of fine-tuning the entire U-Net-based diffusion model, ControlNet freezes the original model, duplicates its encoding layers, and finetunes the copies to learn additional control conditions. 
IP-Adapter\cite{ip-adapter} enhances controllability by introducing decoupled cross-attention layers, which process image features separately from text. This method finetunes fewer parameters and leverages a pretrained image encoder for feature extraction.
In the audio domain, Music ControlNet \cite{music-controlnet} extends ControlNet to the music domain, enabling time-varying control in a pretrained text-to-music diffusion model through three distinct inputs: chromagram for melody, frame-wise energy for dynamics, and beat/downbeat logits for rhythm. 
Guo et al. \cite{multicon_diff} proposed Fusion-Net, which incorporates fine-grained temporal inputs—such as timestamps, pitch contour, and energy contour (similar to RMS)—into a pretrained Text-to-Audio (TTA) model. Their method applies convolutions, linear projections, and self-attention layers.

Recent dual-latent approach for video-to-sound generation used onset timestamps of sounds alongside text prompts to align generated audio with input video. Syncfusion~\cite{syncfusion} trained a timestamp detection network to classify each video frame via supervised learning. An LDM then takes the predicted onset and text as conditions for audio generation.
Although timestamp could be a simple and intuitive feature to control, it requires costly human annotations, which are often ambiguous in defining precise time boundaries.
 
Our Video-Foley framework leverages RMS as an annotation-free feature to bridge video and audio more effectively. Section \ref{ssec:rms} discusses its advantages over timestamp-based features. Video-Foley also uses a ControlNet variant (RMS-ControlNet) that focuses on controlling frame-level intensity using RMS. This approach is similar to T-Foley, but extends it by moving beyond finite sound classes to unconstrained semantic prompts and does not require training from scratch, as it leverages a pretrained TTA model.
As a concurrent work, SonicVisionLM \cite{sonicvisionlm} integrates a ControlNet-based module into a pretrained TTA model to inject onset and offset timestamp information predicted from video. Similarly, ReWaS \cite{rewas} employs a smoothed frequency-mean energy derived from a mel spectrogram, essentially a scaled version of RMS. However, ReWaS differs from our approach in its target data domain (in-the-wild videos vs. clean sound-effect videos) and implementation details (e.g., shorter generation length, lower frame rate).

\section{Proposed Method}
\subsection{Overview}

Fig. \ref{fig:model} illustrates our proposed neural video-to-sound generation model, Video-Foley. It consists of two parts: Video2RMS and RMS2Sound. 
The overall pipeline operates as follows: first, the model takes a video and a prompt as inputs. The prompt, which describes the desired sound, can be either an audio sample or a text description, corresponding to the semantic timbre condition in Fig. \ref{fig:model_type}. Next, Video2RMS predicts the temporal feature of the audio from video, capturing its temporal dynamics. The output serves as the temporal event condition defined in Fig. \ref{fig:model_type}. Finally, RMS2Sound takes both the semantic and temporal conditions and generates the corresponding sound through a diffusion process.

The proposed pipeline effectively processes semantic and temporal information via the dual-latent spaces. By leveraging a video-predictable temporal feature and a semantic prompt, Video-Foley enables video-to-sound generation without costly human annotations or heavy end-to-end training. Unlike single-latent models, Video-Foley’s temporal feature ensures high temporal synchronization, even at commercial-level video frame rates such as 30 fps. Compared to other dual-latent models, our temporal feature is directly derived from the audio waveform while serving as an intuitive condition for audio generation.
Furthermore, our model efficiently generates audio from these two conditions using RMS-ControlNet. Although semantic and temporal attributes are treated separately, they remain interdependent—e.g., hitting glass harder produces a sharper sound, while a dog barking with its mouth wide open results in a louder, more resonant sound. RMS-ControlNet integrates the temporal event condition into the generation process while leveraging a pretrained TTA model for semantic prompt understanding. This approach enables efficient training with audio-only data while ensuring high-fidelity audio generation that aligns with both semantic and temporal conditions.

\begin{figure*}[t]
    \centering
    \resizebox{1\textwidth}{!}{%
        \includegraphics{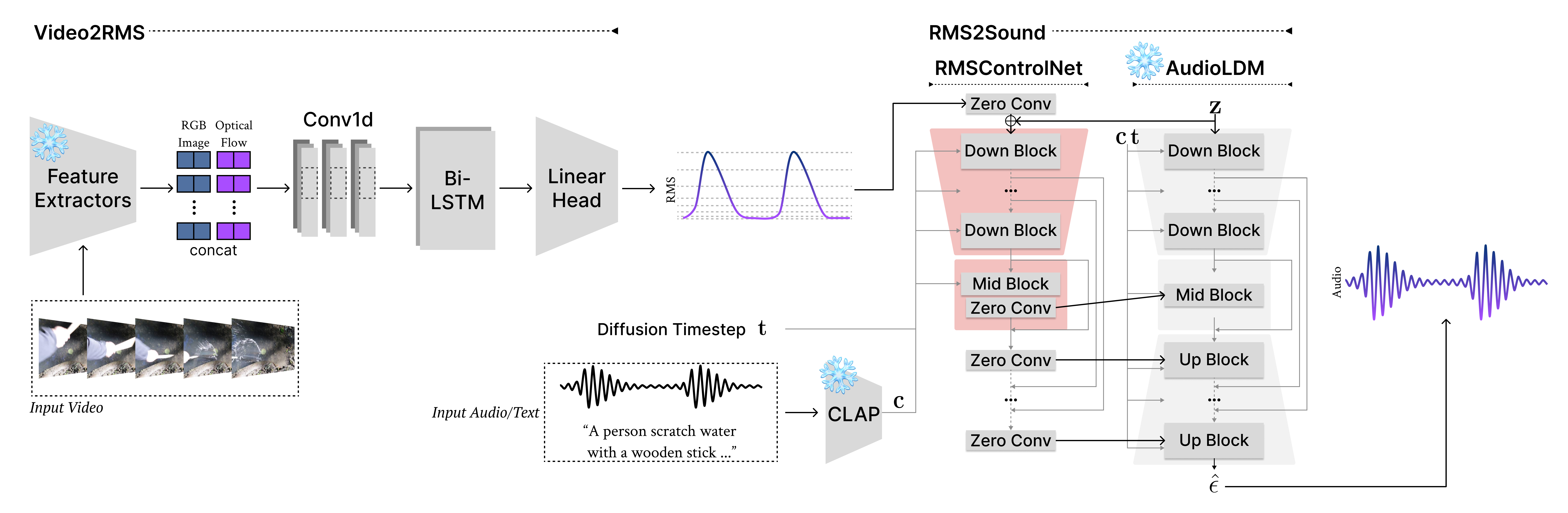}
    }
    \vspace{-0.6cm}
    \caption{Architecture of Video-Foley (Video2RMS and RMS2Sound)}
    \label{fig:video2rms2sound}
    \vspace{-0.3cm}
\end{figure*}

\begin{figure}[t]
    \centering
    \resizebox{0.79\columnwidth}{!}{
        \includegraphics{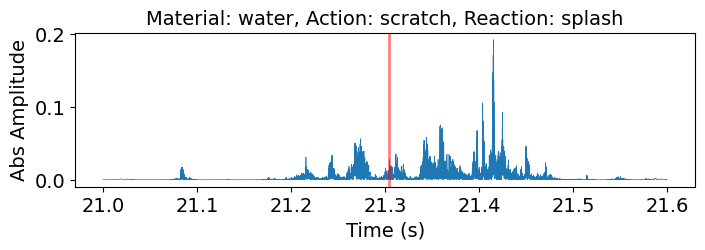}
    }
    \resizebox{0.8\columnwidth}{!}{
        \includegraphics{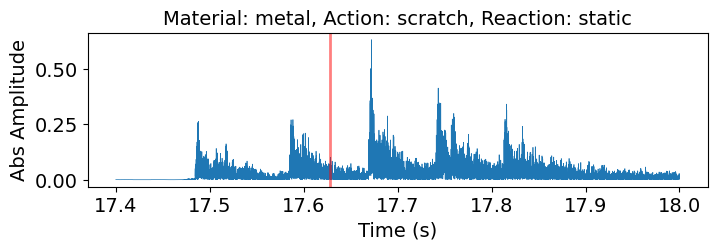}
    }
    \vspace{-0.3cm}
    \caption{Onset annotation examples (red vertical line) with absolute amplitude of the waveform (blue curve) in \textit{Greatest Hits} dataset.}
    \vspace{-0.3cm}
    \label{fig:onset_annot}
\end{figure}

\subsection{RMS as a Temporal Event Condition}
\label{ssec:rms}

We propose to use RMS over timestamps as a temporal feature to shape the audio in alignment with video. RMS serves as the temporal event condition in RMS-ControlNet, guiding the audio generation process.
Here, RMS serves as an effective intermediate feature to shape audio temporally corresponding to video.
We define RMS $R(x)=[R_1,...,R_i,...]$ as a frame-level amplitude envelope feature of audio waveform defined as follows: for the i-th frame,  
\begin{equation}
    R_i(x)=\sqrt{{1\over \omega}\Sigma_{t=ih}^{ih+\omega}x^2(t)}
\label{eq.event_feature}
\end{equation}
where $x(t)\ (t\in[0,T])$ is the audio waveform, $\omega$ is a window size and $h$ is a hop size.

There are two main reasons for selecting RMS. First, timestamps capture only the start and end points of sound events, overlooking crucial audio characteristics such as volume dynamics—for instance, the gradual intensity shift of a moving car, which is difficult to describe in text. RMS effectively represents these aspects, from transient event-based sounds to continuous ambient sounds~\cite{t-foley}, and has already been shown to serve as an effective temporal condition for audio generation, as discussed in Section \ref{ssec:con_audio_gen}. Additionally, human annotation for timestamp is costly and highly subjective, as further discussed in the next paragraph. Moreover, Heller et al. \cite{hybrid_foley} demonstrated that listeners perceive hybrid sounds—Foley sounds combined with the temporal envelope of real recordings—as more realistic than either Foley or real recordings alone. This finding underscores the importance of intensity dynamics in sound generation.


Onset annotation is inherently subjective and lacks a systematic definition, making it prone to inconsistencies. For certain sound events—such as scratching sounds with multiple adjacent attacks or sounds with slow gradual attacks including water, wind, and musical instrument sound—defining an exact onset timestamp is challenging, as it may not align with both the waveform envelope and human perception. In other words, the moment a sound begins physically does not always match how humans perceive its onset \cite{vos1981perceptual}. Fig. \ref{fig:onset_annot} demonstrates examples from the \textit{Greatest Hits} dataset \cite{greatesthits}, highlighting the subjectivity of onset annotations. This subjectivity is critical in video-to-sound generation, where precise temporal synchrony is essential. Inaccurate timestamp annotations can degrade model performance and make timestamp-based evaluations unreliable.

\subsection{Video2RMS} \label{ssec:video2rms}
Video2RMS aims to predict the RMS curve, representing the windowed root mean of squared audio amplitude proportional to intensity, from a sequence of video frames. 
We introduce two key ideas to tackle this problem. First, we propose to discretize the RMS target and formulate the problem as a classification task. 
Since non-ambient action-based sounds are transient and sparse, much of the audio remains nearly silent. Our ablation study showed that training with the L2 loss as a regression task led to poor results, as the model tended to predict silence to reach a local minimum (Fig. \ref{fig:plot_rms}). We discretized the continuous RMS curve into equidistant bins after scaling with the $\mu$-law encoding \cite{wavenet}, formulated as follows:
\begin{equation}
    f(r) = {ln(1+\mu|r|)\over ln(1+\mu)}
\label{eq:mu_law}
\end{equation}
where $r\in[0,1]$ is the RMS value and $\mu+1$ is the number of discretized bins. 
Second, we use the label smoothing to mitigate the penalty for near-correct predictions.
We adopted the Gaussian Label Smoothing (GLS), frequently used in pitch estimation \cite{kum2019joint,crepe}. 
The smoothed label $y$ is formulated as follows:
\begin{align}
    y(k) =
    \begin{cases}
        \exp(-{(c_k-c_{gt})^2\over 2\sigma^2}) & \text{if } |c_k-c_{gt}|\leq W\ (c_k,c_{gt}\neq0)\\
        0 & \text{otherwise}
    \end{cases}
\label{eq:label_smoothing}    
\end{align}
where $k$ is the class index, $c_{gt}$ is the ground-truth class, $\sigma=1$, and $W$ is the smoothing window size determined by the ablation study.

As illustrated in Fig. \ref{fig:video2rms2sound}, the Video2RMS model consists of three 1D-convolutional blocks, two Bi-LSTM layers, and a linear projection head. The architecture is inspired by the visual encoder of RegNet \cite{regnet}, with the key difference that our model includes a linear head to predict the RMS curve using a classification loss, whereas RegNet uses the LSTM output as a temporal video feature implicitly trained with GAN loss.
For input, the BN-Inception network \cite{bn_inception}, pretrained on ImageNet classification\footnote{\url{https://yjxiong.blob.core.windows.net/models/bn_inception-9f5701afb96c8044.pth}}, extracts video features frame-wise from RGB images and 2-channel optical flows. 
For optical flow extraction, pretrained RAFT (\path{Raft_Large_Weights.C_T_SKHT_V2}) in pytorch was used.\footnote{
\url{https://pytorch.org/vision/main/models/generated/torchvision.models.optical_flow.raft_large.html}}
Since BN-Inception is originally designed for 3-channel image inputs, the first convolutional kernel is inflated by averaging across the three channels and duplicating across two axes to accommodate the two-channel optical flow. The feature is taken after the last average pooling layer of the frozen BN-Inception.
The two features are then concatenated for each time frame.
Three convolutional blocks process the local information of the feature sequence.
Each convolutional block includes a convolution layer, a batch normalization layer, and a ReLU activation layer.
Two layers of bidirectional LSTM encode the global information of the features across the time axis. 
Finally, the linear head projects the feature sequence to predict the classification probability for each RMS bin.
The loss function is defined as $L=\sum_{i}CE(\hat{c}_i,c_i)$ where $c_i$ denotes the discretized RMS class label at i-th frame, $\hat{\cdot}$ is the prediction, and $CE$ is the cross entropy loss.

\subsection{RMS2Sound}
To guide the audio generation that reflects both semantic and temporal conditions, we propose RMS2Sound which is a combination of RMS-ControlNet and a frozen TTA model, that generates audio from input RMS and audio-text joint embedding as shown in Fig. \ref{fig:video2rms2sound}. RMS-ControlNet consists of a trainable copy of the encoding layers and the middle block of the backbone TTA model, connected to the frozen backbone layer-wise through zero-initialized convolutional layers. AudioLDM \cite{audioldm}, conditioned on CLAP \cite{clap_ms} embeddings, was used as the backbone TTA model. RMS-ControlNet receives the same input as AudioLDM, except that the noisy latent is summed with the RMS condition. To match the feature dimensions of the RMS condition to those of the noisy latent, we apply a 2D zero-initialized convolutional layer. RMS-ControlNet is trained following the same procedure as the original ControlNet \cite{controlnet}. 
The training loss function is as follows:
\begin{equation}
    \mathbb{E}_{x,t,\epsilon}  \lVert \epsilon - f(z_t,t,C(x),R(x)) \rVert_{2}^{2}
\end{equation}
where $\epsilon$ is the noise injected during forward diffusion process, $x$ is the audio waveform, $z$ is a latent representation of $x$ encoded with a pretrained variational autoencoder, $z_t$ is $z$ at $t$ diffusion timestep, $C$ is the CLAP encoder, and $R$ is the RMS computation. We freeze the parameters of AudioLDM and update only those of RMS-ControlNet. RMS-ControlNet is trained on audio-only data to take advantage of its larger scale compared to video datasets. Since CLAP provides a joint audio-text representation space, RMS2Sound is capable of generating audio from either text or audio prompts. 

\section{Experiments}

\subsection{Dataset}
We used the Greatest Hits dataset\cite{greatesthits} with its official train-test split for training and evaluation. The dataset contains 977 videos of a person making sounds with a wooden drumstick on 17 different materials (wood, metal, rock, leaf, plastic, cloth, water, etc.) using two types of actions (hit, scratch). 
We segmented the videos with denoised audio into 10-second clips without overlap, and resampled to 16kHz for audio and 30fps for video. Each video frame was resized to 344$\times$256 pixels. This resulted in 2,212 training videos (6.14 hours) and 732 test videos (2.03 hours). The training set was used to train Video2RMS, and the test set was used to evaluate both Video2RMS and the entire Video-Foley model. 
To increase extensibility and applicability, we trained RMS-ControlNet using audio-only data from a variety of sounds, rather than limiting it to hit and scratch sounds. We used the FreeSound dataset\cite{wavcaps}, which contains about 6,000 hours of audio. All audio was resampled to 16 kHz.

\subsection{Experimental Details}

\subsubsection{Training} The Video2RMS and RMS2Sound models are trained separately but combined during inference. 
For Video2RMS, RMS was calculated from the audio waveform with a 512 window size and a 128 hop length, following the configuration in T-Foley\cite{t-foley}. By padding $(512-128)/2$ values at both ends of the waveform in reflect mode, we obtained 1250 frames. Then, the RMS was discretized into 64 bins ($\simeq$0.5dB granularity), and Gaussian label smoothing was applied ($W=2)$. The model was trained for 500 epochs using a StepLR scheduler (rate $1e$-$3$, step size 100), with a batch size of 512 using Adam optimizer.
For RMS2Sound, the window and hop length of RMS are set to 1024 and 160 respectively, following AudioLDM's detail. By padding $(1024-160)/2$ values at both ends of the waveform in reflect mode, we obtained 1024 frames. When using the predicted RMS from Video2RMS, nearest-neighbor interpolation is applied to match the feature length. Note that the RMS was not discretized in RMS-ControlNet, i.e. the continuous-valued RMS is the conditioning input.
We initialized AudioLDM using the official checkpoint \texttt{`audioldm-s-full'}\footnote{\url{https://huggingface.co/haoheliu/AudioLDM-S-Full}}. For ControlNet, we used only the weights of the encoder and middle block of the U-Net in the same checkpoint.
RMS-ControlNet was trained for 300k steps using the AdamW optimizer. To maintain training consistency, we adhere to the original AudioLDM configuration (e.g., audio normalization).
The learning rate started at $1e$-$4$ and was halved every 10k steps. During training, only the parameters of RMS-ControlNet were updated.

\subsubsection{Inference} The generated audio duration is 10.24 seconds. For RMS2Sound, we only use Classifier-Free Guidance (CFG) for semantic prompting and do not apply it to RMS conditions, as we did not observe meaningful performance improvements. The CFG is formulated as follows:
\begin{equation}
    \begin{split}
    &\hat{f}(z_t,t,C(x),R(x)) \\
    &\quad \quad \quad = \omega f(z_t,t,C(x),R(x)) + (1-\omega)f(z_t,t,R(x)) 
    \end{split}
\end{equation}
where $\omega$ is a guidance scale, $z$ is a latent representation of audio $x$ encoded with a variational autoencoder (VAE), $z_t$ is $z$ with $t$ times noise added, $C$ is the CLAP encoder, and $R$ is the RMS calculation. Note that a learned null embedding is used instead when CLAP embedding $C(\cdot)$ is not given to the model $f$. 
In our experiment, $\omega$ is fixed to 3.5.

\subsection{Baseline Models}
\label{ssec:baseline}
For comparison with our model, we include two primary baselines for video-to-sound generation: 
CondFoleyGen~\cite{condfoleygen}, which uses audio-visual prompt, and SyncFusion~\cite{syncfusion}, which leverages onset timestamps. Both, like our model, fall under the dual-latent category described in Section \ref{sec:v2s_gen}.
These baselines help validate our choice of RMS as a bridging temporal feature.
All models were trained on Greatest Hits ($\sim$6 hours). 
We also include additional single-latent baselines, 
SpecVQGAN~\cite{specvqgan} and Diff-Foley~\cite{diff-foley}. 
Note that these models do not receive any text or audio as semantic prompts, putting them at a disadvantage in terms of semantic alignment. SpecVQGAN was trained on VGGSound\cite{vggsound} ($\sim$0.4k hours), Diff-Foley was trained on VGGSound and a subset of AudioSet\cite{audioset} ($\sim$1.1k hours). 
Despite the larger size, these in-the-wild datasets sourced from Youtube suffer from low audio-visual quality, duplicates, and visually non-indicative sounds (i.e., weak alignment between visual content and sound such as off-screen sound or background noise~\cite{moviegen,vintage}). 
While acknowledging that a perfectly fair comparison is difficult, we include these representative single-latent models to provide broader context in both objective and subjective evaluation.

All baseline inferences were conducted using official code and checkpoints. Unless otherwise noted, we followed the default configuration choices for inference.
Although CondFoleyGen generates 2 seconds of audio from 15 fps video, the official code was implemented to generate multiples of 2 seconds of audio by adjusting the parameter $W_{scale}$. We set $W_{scale}$ to 5 to generate 10 seconds of audio. 
SyncFusion was trained to generate 5.46 seconds of audio from 15fps video. 
We generated 5-second audio clips and concatenated them. 
For text prompts, we used the same text as Video-Foley. 
SpecVQGAN generates 10 seconds of audio from 21.5 fps video. 
The model `2021-07-30T21-34-25\_vggsound\_transformer' was used.
As Diff-Foley generates 8-second audio from 4 fps video, we made two inferences: one with video frames from 0-8 seconds and another from 2-10 seconds. We then concatenated the entire first segment with the latter 2 seconds of the second segment to produce a 10-second audio. 

For Video2RMS prediction, we provide the model trained in a regression setting, mentioned in Section \ref{ssec:video2rms}, as a baseline. This model is trained with L2 loss $L=\sum_{i}||\hat{R}_i - R_i||^2$ where $R_i$ denotes the continuous RMS value of i-th frame, $\hat{\cdot}$ is the prediction. Other details, such as model architecture or configurations, are the same as in our proposed classification model.

\subsection{Evaluation}
\label{sec:evaluation}
To measure the performance of synchronized video-to-sound generation, three main aspects are considered.
\textit{Semantic Alignment} evaluates how well the timbre and nuance of sound match the material and action type in the video, \textit{Temporal Alignment} examines the accuracy of the start and end timing of a sound event as well as its intensity changes over time, and \textit{Audio Quality} assesses the overall quality of the audio. Both objective and subjective evaluations are conducted. To match our experiment settings, we resampled generated audios to 16 kHz and combined them with the 30 fps videos to create 10 sec video-audio pairs. 

RMS-ControlNet, based on AudioLDM, can use either an audio prompt or a text prompt for timbre conditions. We conducted ablation studies to compare these two prompt methods. For the audio prompt, we simply used ground-truth audio. For the text prompt, we utilized a prompt template: \textit{``A person $\{$action$\}$ $\{$material$\}$ with a wooden stick."} and annotations on material and actions from the Greatest Hits dataset. If there were multiple actions or materials, we made multiple sentences and combined them with \textit{``After that,"}. If no annotation was available, we used \textit{``A person hit something with a wooden stick."} as the default text prompt.

\subsubsection{Objective Evaluation}
To measure overall audio quality, Frechet Audio Distance (FAD) \cite{fad} was used, which is a set-wise distance of audios in embedding space. 
When reference set embeddings $r$ and a generated set embeddings $g$ are given, we calculate the FAD as follows:
\begin{equation}
    \text{FAD}(r, g) = \left\| \mu_r - \mu_g \right\|_2 + \text{tr} \left( \Sigma_r + \Sigma_g - 2\sqrt{\Sigma_r \Sigma_g} \right)
\label{eq:FAD}
\end{equation}
where $\mu_x$ and $\Sigma_x$ are the mean and covariance matrix of the distribution $x$. 
Given that FAD correlation with human perception is embedding-dependent \cite{fad_audio}, we used pretrained PANNs wavegram-log-mel \cite{panns} and CLAP from Microsoft \cite{clap_ms} to extract embeddings through \textit{fadtk}\footnote{\url{https://github.com/DCASE2024-Task7-Sound-Scene-Synthesis/fadtk}}. 

To measure the semantic alignment between audio and video, FAVD \cite{peavs} was used, which is the Frechet distance of concatenated video and audio embeddings. Pretrained VGGish \cite{vggish} and I3D \cite{i3d} were used for audio and video embeddings, respectively. 

\begin{table}[t]
\caption{Performance of Video2RMS module. RMS prediction accuracy (RMS Pred. Acc) is calculated with $\pm$1/3/6 dB tolerance ($\pm$2/5/8 adjacent classes). l.b.: lower bound, disc. RMS (g.t.): discretized version of ground-truth RMS.}
\label{tab:video2rms}
\centering
\resizebox{0.87\columnwidth}{!}{%
\begin{tabular}{lllll}
\toprule
\multirow{2}{*}{Model} &
  \multicolumn{1}{c}{\multirow{2}{*}{E-L1 $\downarrow$}} &
  \multicolumn{3}{c}{RMS Pred. Acc $\uparrow$} \\ \cmidrule(l){3-5} 
 &
  \multicolumn{1}{c}{} &
  \multicolumn{1}{c}{$\pm$1dB} &
  \multicolumn{1}{c}{$\pm$3dB} &
  \multicolumn{1}{c}{$\pm$6dB} \\
\midrule 
random choice (l.b.)           & 0.299  & 0.077 & 0.165 & 0.248 \\ 
\midrule
Regression (baseline) & 0.119 & 0.126 & 0.229 & 0.285 \\
Classification (Ours)             & 0.082  & 0.164 & 0.349 & 0.498 \\
\ \ \ w/ label smoothing & \textbf{0.080}                                        & \textbf{0.165}             & \textbf{0.361}             & \textbf{0.506}             \\
\midrule
disc. RMS (g.t.)               & 0.018 & 1.000  & 1.000 & 1.000 \\
\bottomrule
\end{tabular}%
}
\end{table}

Additionally, the CLAP \cite{clap_laion} score was calculated by measuring the cosine distance between the generated and ground-truth audio pairs in the joint text-audio embedding space.\footnote{Note that this CLAP model is from LAION~\cite{clap_laion}, which differs from the Microsoft model~\cite{clap_ms} used in AudioLDM.} First, we extract embeddings from ground-truth $e$ and generated audio $\hat{e}$ in the audio-text joint embedding space of CLAP. Then, the cosine distance between the two embedding vectors $\cos(e,\hat{e})$ is measured.

Lastly, we used E-L1(Event-L1), the L1 distance between the continuous RMS values of the generated and ground-truth audio as proposed in T-Foley \cite{t-foley}, to measure the temporal synchrony of audio and video. It is defined as the following: 
\begin{equation}
    E\text{-}L1 = {1\over k}\displaystyle\Sigma_{i=1}^k ||E_i-\hat E_i||
\end{equation}
where $E_i$ is the ground-truth event feature of $i$-th frame, and $\hat E_i$ is the predicted one. In this paper, RMS scaled with $\mu$-law encoding is the temporal event feature. For evaluating Video2RMS, E-L1 between the predicted RMS and the ground-truth RMS is measured. In the case of Video-Foley, E-L1 between the RMS extracted from generated audio and ground-truth audio are considered.

\begin{figure}[t]
    \centering
    \resizebox{0.85\columnwidth}{!}{%
        \includegraphics{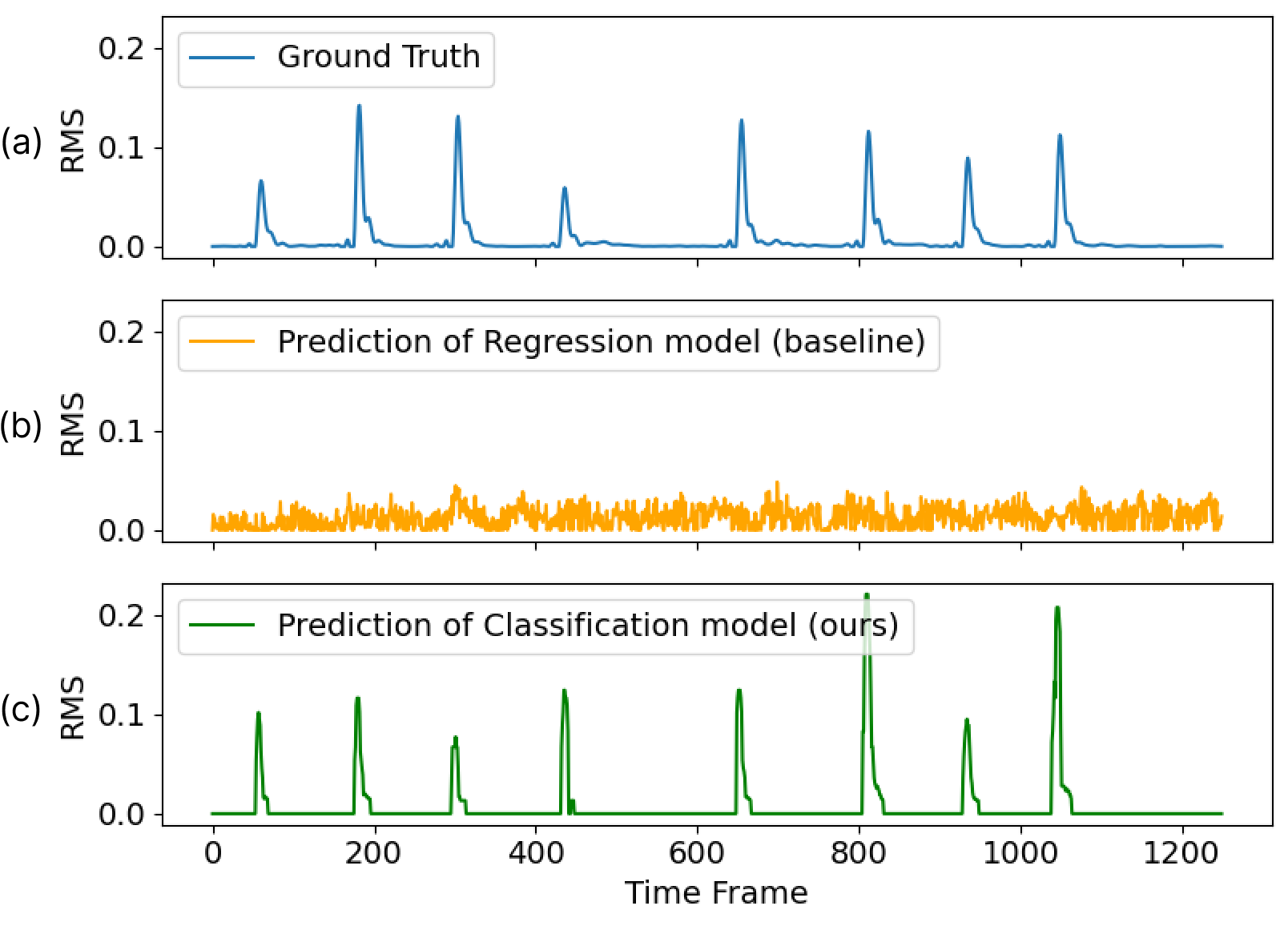}
    }%
    \vspace{-0.3cm}
    \caption{Comparison of the ground-truth RMS curve (a) with the predicted curves from the regression baseline (b) and our classification model (c).}
    \label{fig:plot_rms}
    \vspace{-0.3cm}
\end{figure}

\begin{figure}[t]
    \centering
    \resizebox{0.8\columnwidth}{!}{
        \includegraphics[width=\linewidth] {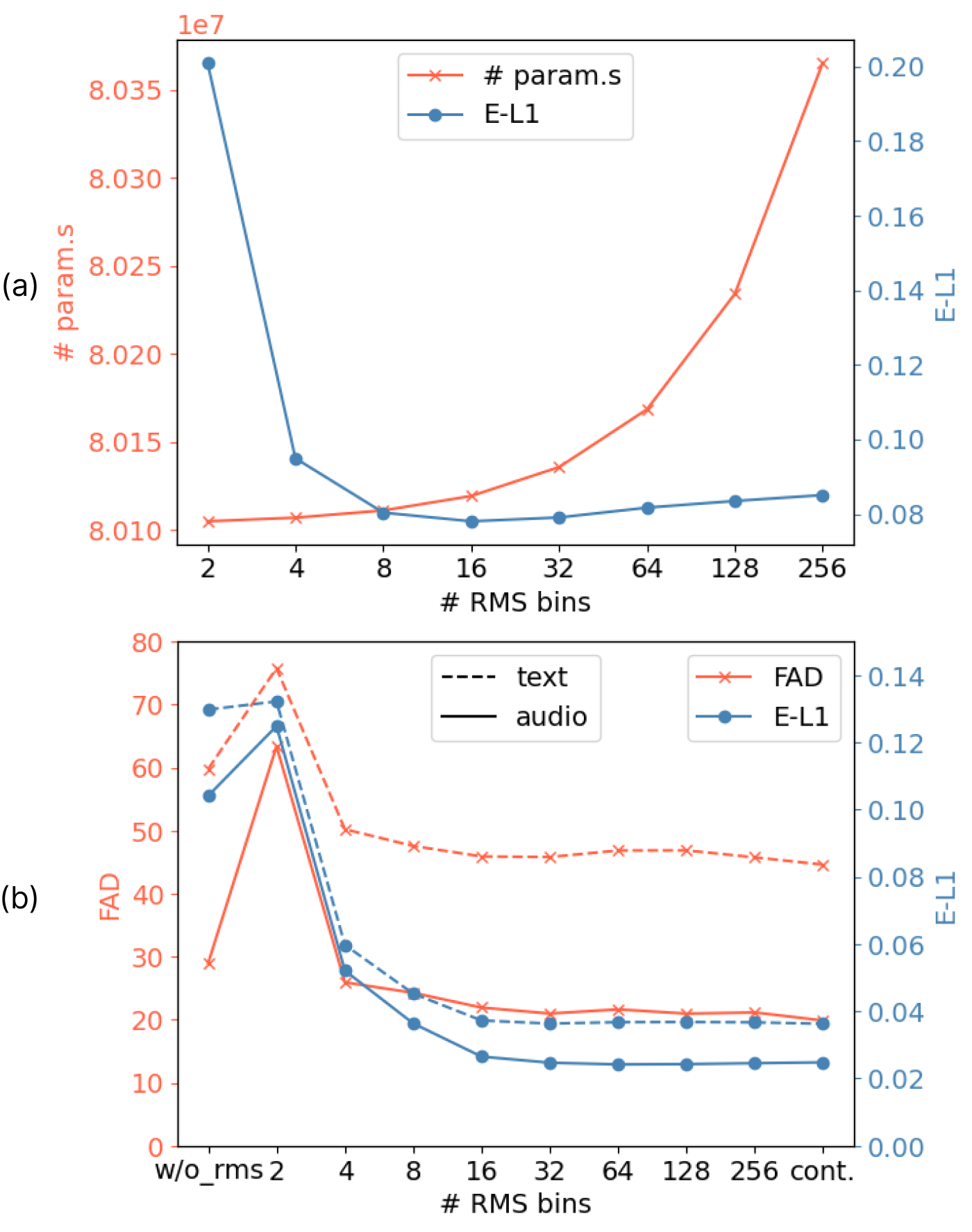} 
    }
    \vspace{-0.2cm}
    \caption{Performance of Video2RMS (a) and RMS2Sound (b) for different numbers of RMS bins. w/o\_rms: without RMS condition (Text-to-Audio\cite{audioldm}), cont.: continuous RMS, no discretization.}
    \label{fig:rms_ablation}
    \vspace{-0.4cm}
\end{figure}

All metrics except FAVD were used to evaluate RMS2Sound, as there is no video input. Classification accuracy with tolerance windows of about $\pm$1/3/6 dB ($\pm$2/5/8 adjacent class bins) measured the RMS prediction performance of Video2RMS, excluding frames where both ground truth and prediction are silent, similar to the previous study \cite{condfoleygen}. This exclusion is necessary because only a small portion of audio frames are non-silent for hit/scratch actions, making the model learn an undesired shortcut for predicting silence and thus failing to effectively capture the performance in non-silent frames.


\subsubsection{Subjective Evaluation}

\begin{table*}[t!]
\caption{Performance of the proposed Video-Foley and other video-to-sound models on \textit{Greatest Hits} testset. $av$: audio-video paired prompt used, $^+$: same CLAP model for train and evaluation. 
}
\label{tab:video2sound}
\centering
{%
\begin{tabular}{@{}rrrrrrrrrr@{}}
\toprule
\multicolumn{1}{l|}{\multirow{2}{*}{Model}} &
  \multicolumn{3}{c|}{Audio Quality} &
  \multicolumn{2}{c|}{Temporal Alignment} &
  \multicolumn{4}{c}{Semantic Alignment} \\
\multicolumn{1}{l|}{} &
  \multicolumn{1}{c}{FAD-P $\downarrow$} &
  \multicolumn{1}{c}{FAD-C $\downarrow$} &
  \multicolumn{1}{c|}{MOS} &
  \multicolumn{1}{c}{E-L1 $\downarrow$} &
  \multicolumn{1}{c|}{MOS} &
  \multicolumn{1}{c}{CLAP $\uparrow$} &
  \multicolumn{1}{c}{FAVD $\downarrow$} &
  \multicolumn{1}{c}{MOS$_{material}$} &
  \multicolumn{1}{c}{MOS$_{action}$} \\ \midrule
Ground Truth       & 0  & 0 & 4.57\scriptsize($\pm$0.08) & 0 & 4.83\scriptsize($\pm$0.06) & 1 & 0 & 4.70\scriptsize($\pm$0.08) & 4.90\scriptsize($\pm$0.04) \\ \midrule
\color{gray}\textit{Audio Prompt} &&&&&&\\
CondFoleyGen$^{av}$\cite{condfoleygen}   
    & 42.2 & 381 & 3.10\scriptsize($\pm$0.13) & 0.148 & 1.93\scriptsize($\pm$0.13) & 0.572 & 1.01 & 2.36\scriptsize($\pm$0.16) & 2.79\scriptsize($\pm$0.17) \\
SyncFusion\cite{syncfusion}
    & 65.9 & 335 & 3.10\scriptsize($\pm$0.13) & 0.150 & 3.10\scriptsize($\pm$0.19) & $^+$0.631 & 4.50 & 3.04\scriptsize($\pm$0.18) & 3.22\scriptsize($\pm$0.19)   \\
Video-Foley (Ours)
    & \textbf{27.2} & \textbf{187} & \textbf{3.93\scriptsize($\pm$0.12)} & \textbf{0.083} & \textbf{4.40\scriptsize($\pm$0.11)} & \textbf{0.644} & \textbf{0.80} & \textbf{3.83\scriptsize($\pm$0.15)} & \textbf{4.56\scriptsize($\pm$0.08)} \\
\color{gray}\textit{Text Prompt}  &&&&&&\\
SyncFusion\cite{syncfusion}
    & 81.6 & 424 & - & 0.162 & - & $^+$0.529 & 5.11 & - & - \\
Video-Foley (Ours)
    & 66.8 & 451 & - & 0.088 & - & 0.476 & 3.28 & - & - \\ \midrule
Text-to-Audio\cite{audioldm}
    & 59.8 & 397 & 2.39\scriptsize($\pm$0.13) & 0.130 & 2.00\scriptsize($\pm$0.13) & 0.443 & 2.67 & 2.78\scriptsize($\pm$0.16) & 3.21\scriptsize($\pm$0.17) \\ \bottomrule
\end{tabular}%
}
\vspace{-0.3cm}
\end{table*}

\begin{figure}[t!]
    \centering
    \resizebox{0.85\columnwidth}{!}{
        \includegraphics{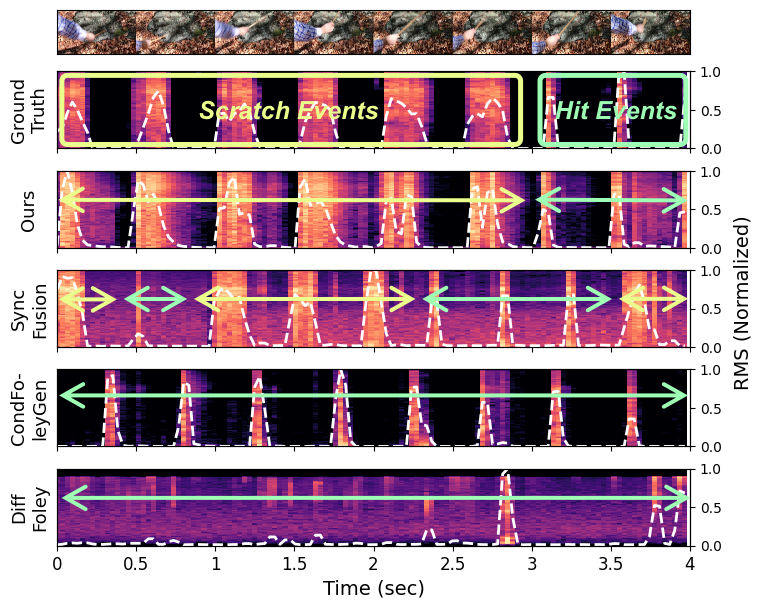}
    }%
    \vspace{-0.3cm}
    \caption{Controlling timbre and energy transition: Video-Foley generates hit and scratch sounds at desired positions using RMS guidance.}
    \label{fig:action_type}
    \vspace{-0.2cm}
\end{figure}

Since a generated sound can be perceptually valid without exactly matching the ground truth~\cite{greatesthits}, we conducted a human listening test to assess the perceptual quality of the generated audio in relation to the input video. A total of 20 participants, including audio ML researchers and audio engineers recruited via email lists and colleagues, were asked to score the audio on a five-point Likert scale based on four criteria: Material / Action / Temporal Alignment, and Audio Quality. Semantic Alignment was divided into two categories to evaluate how well the sound matches the material type and action nuance of the sound events in the video. 
We provided guidelines and video examples to clearly distinguish \textit{Material} and \textit{Action Alignment} from \textit{Temporal Alignment} during evaluation.
The evaluation survey consisted of 12 questions covering different material-action types. Specifically, we excluded ‘None’ from the dataset’s 18 material categories and selected six ‘\{material\}-scratch’ cases where the sound characteristics significantly change by scratching actions. These cases include \textit{plastic-scratch}, \textit{rock-scratch}, \textit{dirt-scratch}, \textit{drywall-scratch}, \textit{gravel-scratch}, and \textit{grass-scratch}. In addition, we selected six ‘\{material\}-hit’ cases from the remaining material categories where the sound characteristics notably change by hitting actions. These cases include \textit{carpet-hit}, \textit{ceramic-hit}, \textit{metal-hit}, \textit{water-hit}, \textit{wood-hit}, and \textit{leaf-hit}. To standardize the length of the sample videos and control evaluator fatigue, we trimmed each video to 4 seconds from the starting point.
Each question presented the ground truth audio and the audio generated by Video-Foley, SyncFusion, Diff-Foley, CondFoleyGen, and AudioLDM in a random order.
Since CondFoleyGen does not support text prompts, audio prompts were used for all models to ensure a fair comparison.
The Mean Opinion Score (MOS) and its 95\% confidence interval were calculated.


\section{Results}

\begin{figure}[t!]
    \centering
    \resizebox{0.85\columnwidth}{!}{
        \includegraphics{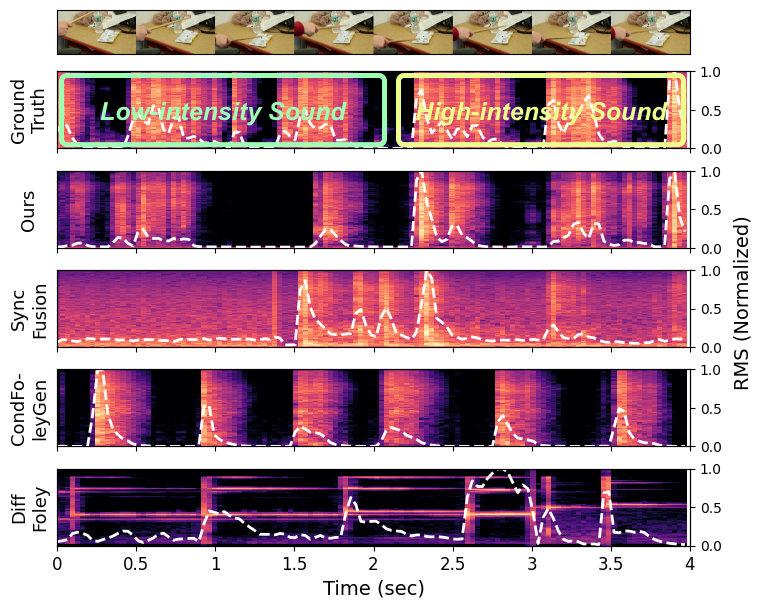}
    }%
    \vspace{-0.3cm}
    \caption{Controlling intensity and nuance: Video-Foley predicts different levels and shapes of the RMS curve for each sound event.}
    \label{fig:intensity}
    \vspace{-0.2cm}
\end{figure}

\subsection{Analysis on Video2RMS}
Table \ref{tab:video2rms} demonstrates the performance of the Video2RMS model.
Our proposed model is compared to random choice (lower bound),  the regression approach, and the scores of discretized ground-truth RMS (upper bound due to information loss).
In the classification setting, the model significantly outperforms the regression baseline across all metrics, validating our decision to approach RMS prediction as a classification problem. As illustrated in Fig. \ref{fig:plot_rms}, the model effectively predicts the RMS curve for sparse audio events, avoiding the shortcut to predict silence as observed in the regression baseline. 
The relatively low accuracy in Acc$\pm$1dB is because it prioritizes predicting a realistic RMS curve over matching the exact magnitude bin.
Finally, the label smoothing improves performance, improving both E-L1 and prediction accuracy.

\subsection{Ablation Study on the Number of RMS Bins}

The number of bins for RMS discretization is a critical parameter that significantly affects both Video2RMS and RMS2Sound. To determine the optimal value, we conducted an ablation study, as presented in Fig. \ref{fig:rms_ablation}. In Video2RMS, we identified a trade-off between prediction performance and computational cost, as shown on the Fig. \ref{fig:rms_ablation}(a); more bins improve temporal synchrony but require higher complexity and more model parameters. As shown in Fig. \ref{fig:rms_ablation}(b), both temporal alignment performance and audio quality in RMS2Sound saturate after bins greater than 64. At 64 bins, we found no performance drop in the quantitative measures when using discretized RMS instead of continuous RMS. 
Therefore, we set the discretization bins to 64. 

\begin{table*}[t!]
\caption{Performance of the proposed Video-Foley and other single-latent video-to-sound models on \textit{Greatest Hits} testset. Regarding train data, \textdagger: \textit{VGGSound}\cite{vggsound} ($\sim$0.4k hr), $\ddagger$: \textit{VGGSound}, subset of \textit{AudioSet}\cite{audioset} ($\sim$1.1k hr), 
otherwise: \textit{Greatest Hits} trainset ($\sim$6 hr).}
\label{tab:video2sound_single}
\centering
{%
\begin{tabular}{@{}rrrrrrrrrr@{}}
\toprule
\multicolumn{1}{l|}{\multirow{2}{*}{Model}} &
  \multicolumn{3}{c|}{Audio Quality} &
  \multicolumn{2}{c|}{Temporal Alignment} &
  \multicolumn{4}{c}{Semantic Alignment} \\
\multicolumn{1}{l|}{} &
  \multicolumn{1}{c}{FAD-P $\downarrow$} &
  \multicolumn{1}{c}{FAD-C $\downarrow$} &
  \multicolumn{1}{c|}{MOS} &
  \multicolumn{1}{c}{E-L1 $\downarrow$} &
  \multicolumn{1}{c|}{MOS} &
  \multicolumn{1}{c}{CLAP $\uparrow$} &
  \multicolumn{1}{c}{FAVD $\downarrow$} &
  \multicolumn{1}{c}{MOS$_{material}$} &
  \multicolumn{1}{c}{MOS$_{action}$} \\ \midrule
Ground Truth       & 0  & 0 & 4.57\scriptsize($\pm$0.08) & 0 & 4.83\scriptsize($\pm$0.06) & 1 & 0 & 4.70\scriptsize($\pm$0.08) & 4.90\scriptsize($\pm$0.04) \\ \midrule
\color{gray}\textit{No Prompt}    &&&&&&\\
SpecVQGAN$^\dagger$\cite{specvqgan}                         
    & 101.0 & 579 & - & 0.148 & - & 0.323 & 6.42 & - & - \\
Diff-Foley$^\ddagger$\cite{diff-foley}              
    & 87.0 & 550 & 2.11\scriptsize($\pm$0.11) & 0.166 & 1.86\scriptsize($\pm$0.14) & 0.403 & 4.61 & 1.78\scriptsize($\pm$0.13) & 2.38\scriptsize($\pm$0.17) \\
\color{gray}\textit{Audio Prompt} &&&&&&\\
Video-Foley (Ours)
    & \textbf{27.2} & \textbf{187} & \textbf{3.93\scriptsize($\pm$0.12)} & \textbf{0.083} & \textbf{4.40\scriptsize($\pm$0.11)} & \textbf{0.644} & \textbf{0.80} & \textbf{3.83\scriptsize($\pm$0.15)} & \textbf{4.56\scriptsize($\pm$0.08)} \\ \bottomrule
\end{tabular}%
}
\vspace{-0.3cm}
\end{table*}

\subsection{Analysis on Video-to-Sound}
\label{ssec:analysis_v2s}

\subsubsection{Quantitative Study} 
Table \ref{tab:video2sound} compares the performance of Video-Foley with other dual-latent baselines on the GreatestHits test set. 
Video-Foley achieved state-of-the-art performance across all objective metrics as well as the human MOS. Notably, it showed a significant performance gap not only in temporal alignment but also in semantic material alignment compared to the audio-visual cued model (CondFoleyGen) and the onset-based model (SyncFusion). This suggests that RMS conditioning is superior for video-to-sound generation, because it conveys both timing and intensity dynamics, providing more detailed information than simple timestamps. Furthermore, this temporal feature can imply the timbre and nuance of the sound through its curve shape, complementing the semantic prompt. Importantly, our model does not require timestamp annotations during training.

In every aspect, including audio quality, Video-Foley also outperforms AudioLDM~\cite{audioldm}, the frozen TTA model in RMS2Sound. This suggests that an additional RMS condition, well matched with the prompt, can help the model generate higher fidelity audio, consistent with the results in Fig. \ref{fig:rms_ablation}.
Video-Foley and SyncFusion, trained exclusively with audio prompts, perform better with audio prompts than text. The complexity of describing multiple sound events over 10 seconds with text versus audio may also contribute to this trend.

Table \ref{tab:video2sound_single} presents the performance comparison with single-latent models to provide additional points of reference for both objective and subjective evaluations. Since these models do not use semantic prompts, they rely solely on video input for semantic alignment, putting them at a disadvantage. Diff-Foley, despite incorporating temporal information for audio-visual joint space learning, lagged in temporal performance. This may be attributed to the limited temporal alignment granularity of its video encoder (4fps) or domain mismatch between its training data and Greatest Hits, which likely led to the generation of visually irrelevant sounds common in noisy in-the-wild datasets as discussed in Section \ref{ssec:baseline}.

\subsubsection{Qualitative Study}
Extensive case studies were conducted to demonstrate the performance and controllability of Video-Foley. Our analysis underscores that the intensity level and energy transition in RMS are often associated with the timbre and nuance of sound, consistent with the findings of the previous study\cite{t-foley}. We plot the mel-spectrogram and normalized RMS of the generated audio from each model.
Fig. \ref{fig:action_type}
shows the synergy of complex prompts with RMS. Only Video-Foley generates hit or scratch sounds at the right moment, as our model can distinguish the timing and type of each sound event from the shape of the RMS curve even for complex audio or text prompts with multiple events. Onset-based models only predict when to make a sound but cannot distinguish different timbres for each event. In contrast, ours can control both the timing and the corresponding timbre by modifying the RMS.
Fig. \ref{fig:intensity} 
illustrates the controllability and high audio-visual alignment of Video-Foley. Only ours effectively predicts and recommends the appropriate RMS level and transition curve, ensuring synchronization with the input video. This includes not only timing but also the intensity and nuance of sound events.
These capabilities are due to Video2RMS's ability to distinguish action types (e.g., hit and scratch), timing, and intensity and predict their corresponding energy transitions, and RMS2Sound's ability to generate appropriate timbre and nuance at the corresponding timings. Additionally, RMS helps enhance temporal alignment. 

\subsection{Ablation Study for Video2RMS}

\subsubsection{Ablation Study on Video Features} 
Table \ref{tab:feature_ablation} shows the objective metric scores of Video2RMS depending on the input video features. The best overall performance was achieved when both RGB and optical flow features were used. Removing either feature led to a performance drop, but excluding the optical flow resulted in a more significant decrease. This suggests that inter-frame differences captured by the optical flow are crucial for predicting temporal audio features like RMS. However, the RGB feature also enhances performance by providing semantic information, such as the presence of sound-related objects in the visual scene.

\subsubsection{Ablation Study on Label Smoothing} 
\begin{figure}[t!]
    \centering
    \resizebox{0.65\columnwidth}{!}{%
        \includegraphics{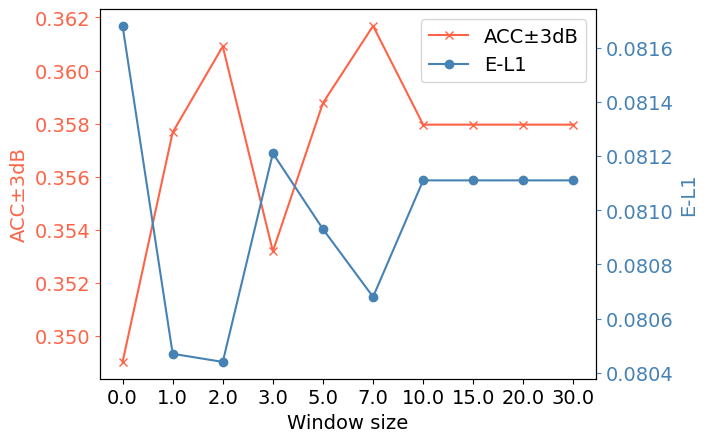}
    }%
    \vspace{-0.2cm}
    \caption{Ablation study on the window size of label smoothing in Video2RMS}
    \label{fig:gls_ablation}
    \vspace{-0.2cm}
\end{figure}

Fig. \ref{fig:gls_ablation} illustrates the performance of Video2RMS with different label smoothing window sizes $W$ in Equation \ref{eq:label_smoothing}. We found that $W=2$ offers the best balance between E-L1 and accuracy. For larger window sizes, the model qualitatively produces more jitter in the RMS curve. The prediction performance saturates after $W=10$, resulting in a spiky RMS curve and a poor overall performance. In addition, using Gaussian label smoothing ($W>0$) consistently improved performance at any window size.

\begin{table}[t!]
\caption{Performance of Video2RMS module on different input video features. OF: Optical Flow, RGB: RGB image.}
\label{tab:feature_ablation}
\centering
\resizebox{0.65\columnwidth}{!}{%
\begin{tabular}{@{}lllll@{}}
\toprule
\multirow{2}{*}{Model} &
  \multicolumn{1}{c}{\multirow{2}{*}{E-L1 $\downarrow$}} &
  \multicolumn{3}{c}{RMS Pred. Acc $\uparrow$} \\ \cmidrule(l){3-5} 
 &
  \multicolumn{1}{c}{} &
  \multicolumn{1}{c}{$\pm$1dB} &
  \multicolumn{1}{c}{$\pm$3dB} &
  \multicolumn{1}{c}{$\pm$6dB} \\ 
\midrule
Video2RMS    & \textbf{0.080} & \textbf{0.165} & \textbf{0.361} & \textbf{0.506}\\ 
\ \ \ w/o RGB & 0.081 & 0.155 & 0.352 & 0.497 \\
\ \ \ w/o OF & 0.088 & 0.149 & 0.335 & 0.470 \\
\bottomrule
\end{tabular}%
}
\end{table}

\subsection{Ablation Study for RMS2Sound}
Table \ref{tab:rms2sound} summarizes the performance of RMS2Sound on audio and text prompts with ground-truth RMS conditions. The discretized RMS (64 bins) performed comparably to the original continuous RMS in terms of audio quality, semantic similarity, and temporal alignment. In contrast, the vanilla TTA model without RMS conditioning (AudioLDM) underperformed in every metric. This supports our assumption that realistic RMS conditions enhance the overall quality of the generated audio.


\subsection{Unlocking the Potential of RMS-ControlNet}
\label{ssec:unlock_rmscontrolnet}

\begin{figure}[t]
    \centering
    \resizebox{0.85\linewidth}{!}{
        \includegraphics{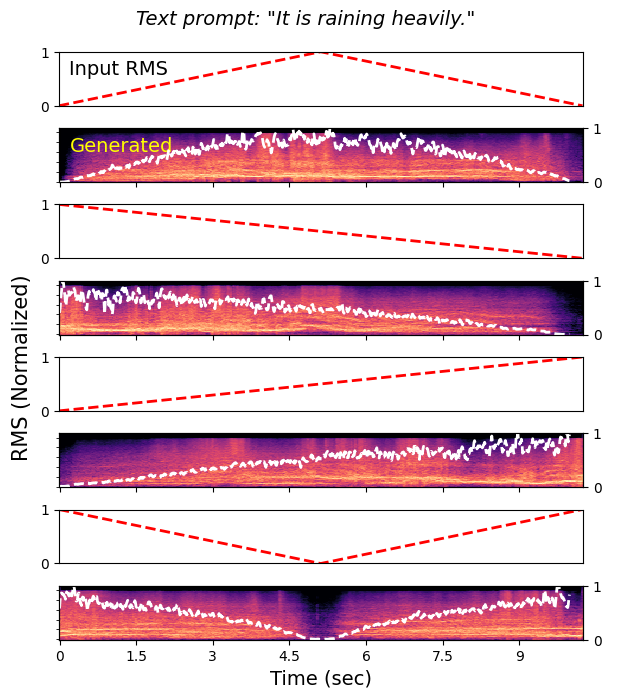}
    }
    \vspace{-0.3cm}
    \caption{
    RMS-ControlNet can control the 
    energy transition
    while reflecting the semantic text prompt.
    }
    \label{fig:rmscontrolnet_rms}
    \vspace{-0.3cm}
\end{figure}
\begin{figure}[t]
    \def\figwidth{0.85\linewidth}
    \centering
    \resizebox{\figwidth}{!}{
        \includegraphics{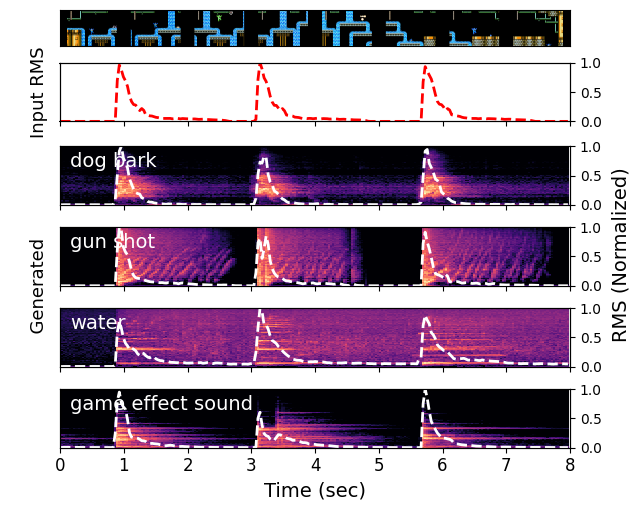}
    }
    \centering
    \resizebox{\figwidth}{!}{
        \includegraphics{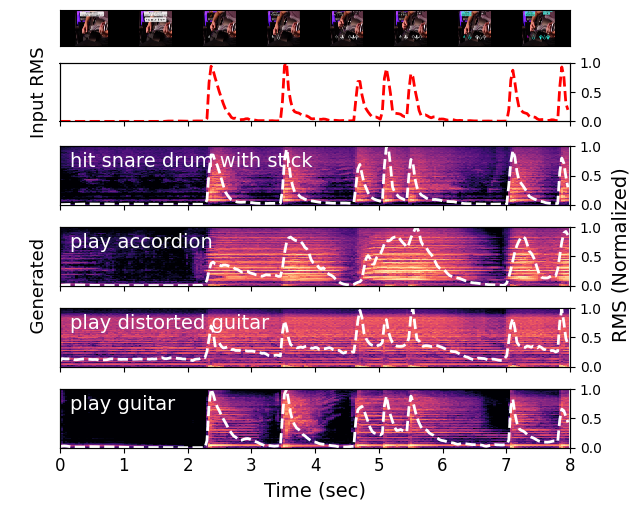}
    }
    \vspace{-0.3cm}
    \caption{
    RMS-ControlNet can control the sound source and nuance through a text prompt while controlling timing and intensity through RMS conditions.\\ }
    \label{fig:rmscontrolnet_text}
    \vspace{-0.3cm}
\end{figure}

RMS-ControlNet, trained for additional temporal event guidance with RMS condition on top of the pretrained TTA model (AudioLDM), shows great potential in controllable audio generation tasks. 
We provide demos to showcase its high controllability, which prior TTA models were not able to achieve.
Fig. \ref{fig:rmscontrolnet_rms} shows how RMS can be simply and intuitively used for temporal guidance.
With the same text prompt, RMS-ControlNet guides AudioLDM to generate audio that matches different input RMS conditions (A-shaped, monotonic decrease, monotonic increase, and V-shaped) that reflect varying distance from the source while maintaining audio semantics (car passing sound). Such intensity dynamics are often used in Foley sound generation, which current text-to-audio models struggle to reflect with sufficient temporal accuracy.
Fig. \ref{fig:rmscontrolnet_text} shows how text prompt can adjust audio semantics along with RMS guidance. While preserving the timing of sound events, users can control various audio semantics such as the sound source and timbre.
This highlights RMS-ControlNet's ability to guarantee high controllability in RMS guidance for timing and intensity while preserving the power in TTA generation. 
Note that these sound sources are not part of GreatestHits; their generation leverages the knowledge embedded in the frozen TTA backbone.

\begin{table}[t!]
\caption{Performance of RMS2Sound module. w/o RMS: pretrained AudioLDM without RMS condition, disc. RMS: discretized RMS in 64 bins, cont. RMS: continuous RMS.}
\label{tab:rms2sound}
\centering
\resizebox{0.7\columnwidth}{!}{%
\begin{tabular}{@{}lllll@{}}
\toprule
Model                             & FAD-P↓ & FAD-C↓ & CLAP↑     & E-L1↓         \\ \midrule
\color{gray}\textit{Audio Prompt} &        &        &               &               \\
w/o RMS\cite{audioldm}              & 29.0  & 194 & 0.619 & 0.104 \\
disc. RMS                     & 21.6  & 154  & \textbf{0.686}  & \textbf{0.024} \\
cont. RMS                         & \textbf{19.9}  & \textbf{152} & 0.657 & 0.025 \\
\color{gray}\textit{Text Prompt}  &        &        &               &               \\
w/o RMS\cite{audioldm}            & 59.8  & 397 & 0.443 & 0.130 \\
disc. RMS                    & 46.8  & 333 & 0.504 & 0.037 \\
cont. RMS                         & \textbf{44.6}  & \textbf{323} & \textbf{0.531} & \textbf{0.036} \\ \bottomrule
\end{tabular}%
}
\end{table}

\section{Discussions}

\subsection{Future Works}
Our Video-Foley model has four main limitations: two stemming from its architecture and two from the Greatest Hits dataset.
Regarding the architecture, our model does not capture temporal dynamics in audio semantics, as CLAP compresses variations in sound sources and timbre into a single aggregated vector \cite{t-clap}. Additionally, the Video2RMS module predicts RMS solely from video input, ignoring audio or text prompts, which limits controllability — users cannot adjust sound timbre temporally or specify particular sound sources. Addressing these issues requires incorporating sequential features to encode temporal changes and integrating prompts into the Video2RMS module for richer semantic guidance.
Greatest Hits dataset consists of mono-sourced audio, preventing the model from handling multiple simultaneous sound sources. Furthermore, all sound sources are in the foreground, restricting spatial awareness. Expanding the dataset to include overlapping sounds and background elements is crucial for improving the model's ability to process complex auditory scenes.

Moreover, we emphasize a broader challenge: the absence of a large-scale, high-quality video dataset that combines precise audio-visual synchrony with curated Foley sound design. While larger datasets such as VGGSound \cite{vggsound} ($\sim$0.4k hours) exist, their suitability for Foley sound generation is limited. As VGGSound is sourced from open-domain platforms like YouTube, it suffers from low audio-visual quality, duplicated content, and visually non-indicative sounds (e.g., off-screen or background noise) \cite{moviegen,vintage}. This underscores the need to construct a high-quality video dataset specifically tailored for Foley applications, with carefully designed and diverse sound categories relevant to multimedia production.

Finally, we observe a lack of standardized metrics for evaluating audio-visual temporal synchrony. While E-L1, a hand-crafted distance metric, effectively captures general quality trends, small improvements in perceptual synchrony may not be reflected strictly in its value. We also considered PEAVS \cite{peavs}, a neural model trained to predict human opinion scores for synchrony. However, PEAVS showed no meaningful correlation with human Mean Opinion Scores (MOS), with a Spearman’s correlation of 0.49 ($p$ = 0.33 $>$ 0.05), compared to E-L1, which achieved a stronger correlation of -0.83 ($p$ = 0.042 $<$ 0.05).
We attribute this discrepancy to two main factors. First, PEAVS has a limited training dataset, using only 200 source videos from the AudioSet~\cite{audioset} evaluation split, augmented to 18.2K samples with artificial distortions. Despite the augmentation, this dataset represents only a small portion of AudioSet, which may reduce generalizability to the entire AudioSet or Greatest Hits. Second, AudioSet videos often contain poor-quality audio-visual pairs with off-screen or irrelevant sounds, potentially leading PEAVS to favor flawed outputs. These findings highlight the urgent need for a more robust and generalizable neural metric for assessing temporal synchrony in video-to-sound generation.

\subsection{Broader Impact}
While our research advances video-to-sound and controllable audio generation, it raises ethical concerns, particularly regarding the potential misuse of realistic audio-visual synthesis. The ability to generate synchronized, high-fidelity sound could contribute to deepfake technology, facilitating misinformation, privacy violations, and the fabrication of deceptive media. Such risks pose serious challenges in media authenticity, public trust, and human rights. To mitigate these threats, it is crucial to establish ethical guidelines and implement safeguards against malicious use
to ensure responsible deployment. Continued scrutiny and proactive governance are essential to balance innovation with societal protection.

\section{Conclusion}
We propose Video-Foley, a two-stage video-to-sound model using RMS as a temporal feature. 
RMS offers three key advantages over timestamps: it does not require human annotation, is closely linked to semantic information, and is easy to control. 
Our quantitative and qualitative studies demonstrate that RMS conditioning enhances both temporal and semantic audio-visual synchrony while ensuring high controllability, thanks to its synergy with audio or text prompts. 
We believe RMS is an effective and intuitive control factor for users, as highlighted in Section \ref{ssec:unlock_rmscontrolnet}. Video2RMS may provide an excellent starting point for creators to refine and shape their desired sound. 
Additionally, the two-stage framework operates without joint training while ensuring high performance. RMS2Sound leverages a pretrained TTA model and benefits from training on large-scale audio-only data, addressing the scarcity of clean, large-scale audio-visual datasets.
We believe our work provides an important initial step towards achieving precise audio-visual temporal synchronization, a critical goal in video-to-sound generation.


\bibliographystyle{IEEEbib}
\bibliography{refs}


\newpage
\begin{IEEEbiography}[{\includegraphics[width=1in,height=1.25in,clip,keepaspectratio]{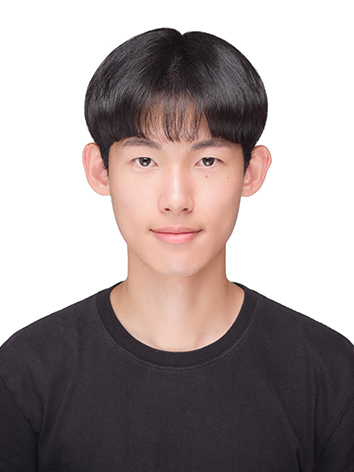}}]{Junwon Lee}
is a PhD student in the Graduate School of Artificial Intelligence at the Korea Advanced Institute of Science and Technology (KAIST) in South Korea. He obtained his M.S. degrees in Artificial Intelligence, and B.S. in Electrical Engineering from Korea Advanced Institute of Science and Technology (KAIST). His research focuses primarily on controllable audio generation and multimodal understanding with deep learning. His research spans across various tasks related to sound effects and music, using text or visuals.
\end{IEEEbiography}

\begin{IEEEbiography}[{\includegraphics[width=1in,height=1.25in,clip,keepaspectratio]{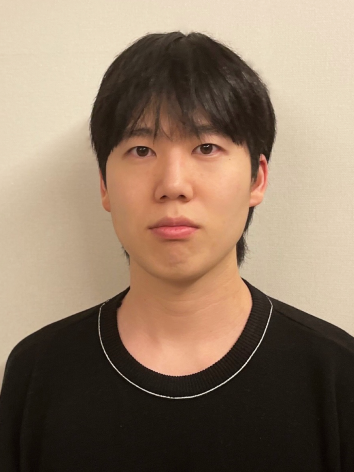}}]{Jaekwon Im}
is a Ph.D. student at the Graduate School of Culture Technology at the Korea Advanced Institute of Science and Technology (KAIST), South Korea. His research focuses on audio enhancement, source separation, and audio generation. He was the organizer of the "Foley Sound Synthesis" challenge at DCASE 2023. He is a reviewer for TASLP and ICASSP.\end{IEEEbiography}

\begin{IEEEbiography}[{\includegraphics[width=1in,height=1.25in,clip,keepaspectratio]{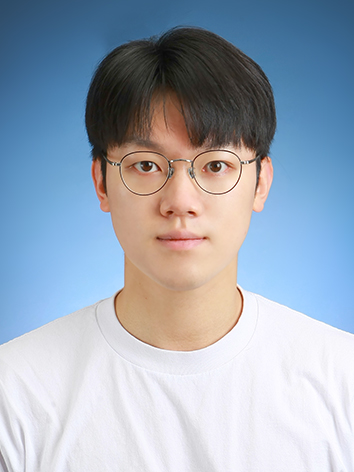}}]{Dabin Kim}
is a master’s student at the Graduate School of Culture Technology, Korea Advanced Institute of Science and Technology (KAIST), South Korea. He received his B.S. degree in Art \& Technology from Sogang University, where he worked on sound synthesis and algorithmic composition systems for multimodal media. His research currently focuses on developing control systems for audio generative models to improve their applicability in real-world sound design and music production.\end{IEEEbiography}

\begin{IEEEbiography}[{\includegraphics[width=1in,height=1.25in,clip,keepaspectratio]{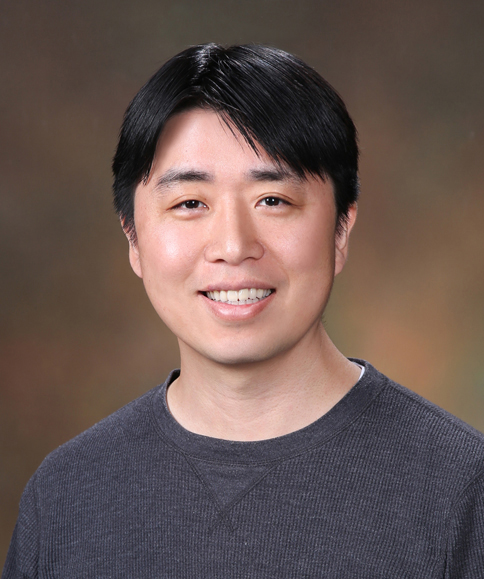}}]{Juhan Nam}
Dr. Juhan Nam is an Associate Professor in the Graduate School of Culture Technology at the Korea Advanced Institute of Science and Technology (KAIST) in South Korea. He received his Ph.D. in Music from Stanford University, where he studied at the Center for Computer Research in Music and Acoustics (CCRMA). Dr. Nam also holds an M.S. degree in Electrical Engineering from Stanford University and a B.S. degree in Electrical Engineering from Seoul National University. His research interests include the application of digital signal processing and machine learning to music and audio. Dr. Nam is a member of the IEEE.\end{IEEEbiography}


\end{document}